\documentclass{ws-procs9x6}
\usepackage{axodraw}

\def\cal{\mathcal}  

\newcommand{\ts}[1]{\mbox{\scriptsize {#1}}}
\newcommand{\tst}[1]{\mbox{\tiny {#1}}}

\newcommand{\half}{\frac{1}{2}}

\newcommand{\spic}[1]{\;\parbox[c]{21pt}{\begin{picture}(21,21)(0,0)
\SetWidth{1.0}\SetScale{0.7} #1 \end{picture}}\;}

\newcommand{\spicc}[1]{\;\parbox[c]{42pt}{\begin{picture}(42,21)(0,0)
\SetWidth{1.0}\SetScale{0.7} #1 \end{picture}}\;}

\begin{document}

\title{$\Phi$-derivable approximations in gauge
theories\footnote{\uppercase{T}alk presented by
\uppercase{A}.~\uppercase{A}rrizabalaga.}}

\author{A.~Arrizabalaga and J.~Smit}

\address{Institute for Theoretical Physics, University of Amsterdam, \\
	Valckenierstraat 65, 1018 XE Amsterdam, The Netherlands.   
	}

\maketitle

\abstracts{We discuss the method of $\Phi$-derivable approximations in gauge
theories. There, two complications arise, namely the violation of Bose
symmetry in correlation functions and the gauge dependence. For the latter we argue that the error introduced by the gauge dependent terms is controlled, therefore not invalidating the method. }

\section{Introduction}
As it is known, calculations in terms of bare quantities usually fail to
describe collective phenomena in high-energy plasmas. A reorganization of
perturbation theory is typically called for. Among the different approaches
proposed, variational methods based on the 2PI effective
action\cite{Baymandcompany} seem to be promising. They have been used to
explore the long time behaviour of non-equilibrium phenomena in large $N$
scalar theories\cite{berges}. In addition,
they have provided quantitative predictions for thermodynamical quantities of
the quark-gluon plasma\cite{blaizot,peshier} which match the lattice results
down to $T\sim 3T_c$.\\
\indent Let us recall briefly how these methods work and why they might be
suited to the description of collective phenomena. They are based on the 2PI
effective action $\Gamma_{\ts{2PI}}$, defined as the Legendre transform
of the generating functional of correlation functions with a bilocal current
introduced. The generating functionals with currents $J$ and $K$ are given by 
\begin{equation}
 Z[J,K]= e^{iW[J,K]}=\int \cal{D} \varphi\, e^{i\left( S[\varphi]+J_i\varphi_i+\half \varphi_i K_{ij} \varphi_j \right)}.
\end{equation}
Hence, the 2PI effective action is defined as 
\begin{equation} 
\Gamma_{\ts{2PI}}[\phi,G] =W[J,K]-J_i\phi_i-\half K_{ij}\left(
\phi_i\phi_j+G_{ij}\right).\\
\label{legendre}
\end{equation}
\indent The indices stand for all field attributes and sum/integration over
repeated ones is understood. The quantities $\phi_i=\delta W /\delta J_i$ and
$G_{jk}=2\,\delta W/\delta K_{jk}-\delta W /\delta J_j \,\delta W /\delta J_k$ are the mean field and the connected 2-point function.  
The effective action $\Gamma_{\ts{2PI}}$ can be written as
\begin{equation}
\Gamma_{\ts{2PI}}[\phi,G]=S_{\ts{free}}[\phi]+\frac{i}{2}\mbox{Tr}\left\{ \log G^{-1}+G(G_{\ts{free}}^{-1}-G^{-1})\right\}-i\Phi[\phi,G]
\end{equation} 
with  
\begin{equation}
\Phi[\phi,G]=\ \parbox[c]{55mm}{
\epsfxsize=55mm
\epsfbox{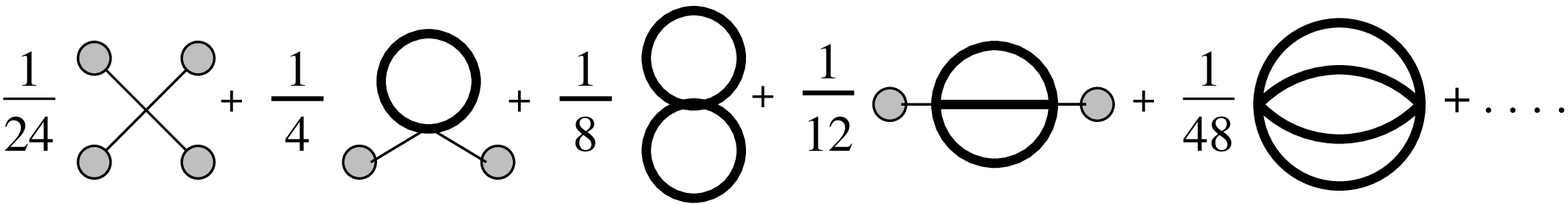}},
\end{equation}
i.e. $\Phi$ is the set of closed skeleton diagrams with full propagators $G$
(thick lines) and mean fields $\phi$ (lollipops). This representation of the
effective action is exact. A truncation of the functional $\Phi$ defines what
is called a \emph{$\Phi$-derivable approximation}. The resulting truncated action
$\Gamma_{\ts{2PI}}^{\ts{tr}}$ is taken to be the fundamental action of the
theory. The corresponding physical mean fields and 2-point functions $\phi_{\ts{ph}}$ and $G_{\ts{ph}}$ are those at the stationary point of $\Gamma_{\ts{2PI}}^{\ts{tr}}$, i.e.
\begin{equation}
\frac{\delta \Gamma_{\ts{2PI}}^{\ts{tr}}}{\delta \phi}\Big|_{\phi_{\ts{ph}},
G_{\ts{ph}}}=0\ ,\ \ \frac{\delta \Gamma_{\ts{2PI}}^{\ts{tr}}}{\delta
G}\Big|_{\phi_{\ts{ph}}, G_{\ts{ph}}}=0.
\label{stationarity}
\end{equation} 
As an example, if we had $\phi=0$ and we took into $\Phi$ only the ``eight'' diagram $\parbox[c]{2.5mm}{
\epsfxsize=2.5mm
\epsfbox{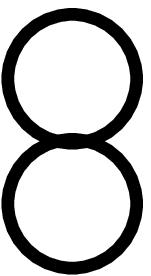}}$, the corresponding self-consistent equation for the physical 2-point function would be
\begin{equation}
 G_{\ts{ph}}^{-1}=G_{\ts{free}}^{-1}+\parbox[c]{8mm}{
\epsfxsize=10mm
\epsfbox{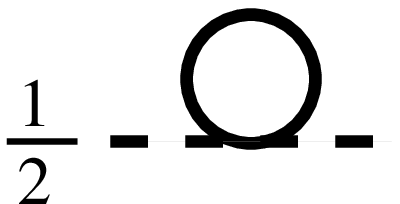}}^{G_{\tst{ph}}}.
\label{example}
\end{equation}

\indent The advantage of using the 2PI effective action is that it depends
explicitely on the exact (dressed) 2-point function. Truncating $\Phi$
according to some ordering principle (coupling constant, $1/N$,...)
defines as in (\ref{example}) an approximate physical 2-point function, which is obtained
self-consistently. Having both the mean field and 2-point function as
dynamical variables allows, for instance, to study a system of unstable
particles or a system with large fluctuations around the mean field (as is the
case out of equilibrium). In addition, $\Phi$-derivable approximations comply with global
symmetries of the hamiltonian (in particular guaranteeing energy-momentum
conservation) and are thermodynamically consistent. These reasons, and the
fact that the method is systematically improvable, make $\Phi$-derivable approximations a promising approach to study high-energy plasmas. \\
\indent Nevertheless, there are still some complications in the method. The truncation of the exact functional $\Phi$ (i.e., with all the diagrams) introduces vertices with the wrong Bose symmetry and, in gauge theories, computed physical quantities might still depend on the gauge condition.
These intricacies constitute the subject of my talk.

\section{Violation of Bose symmetry in correlation functions}
Imagine we were interested in obtaining a $n$-point correlation function ($n>2$) as defined within a $\Phi$-derivable approximation. In the case of the exact theory we could take Eq.~(\ref{legendre}) and expand both $\Gamma_{\ts{2PI}}[\phi,G]$ and $W[J,K]$ in their arguments. We could also formally expand $\phi[J,K]$ and $G[J,K]$ in the currents $J$ and $K$. Identifying a given order in powers of $J$ and $K$ in both sides of Eq.~(\ref{legendre}) would produce coupled equations between various derivatives $\delta_\phi \delta_G \cdots \Gamma_{\ts{2PI}}$ and $\delta_J \delta_K \cdots W$. The later are related to connected correlation functions $\langle \varphi\cdots\varphi\rangle_{c}$. Solving the set of equations one obtains various correlation functions in terms of derivatives of $\Gamma_{\ts{2PI}}$. Solving the same set of equations but with $\Gamma^{\ts{tr}}_{\ts{2PI}}$ would give the correlation functions for the $\Phi$-derivable approximation. \\
\indent Consider for example the scalar $\lambda \phi^4$ theory in the
symmetric phase. Following the procedure above one obtains for the 4-point function $G^{(4,\ts{tr})}$
the following equation 
\begin{equation}
\frac{\delta^2 \Gamma^{\ts{tr}}_{\ts{2PI}}}{\delta G_{ij}\delta G_{mn}}\Big( G^{(4,\ts{tr})}_{mnkl}+G_{mk}G_{nl}+G_{ml}G_{nk}\Big)=i(\delta_{ik}\delta_{jl}+\delta_{il}\delta_{jk})
\label{4-point}
\end{equation} 
In the $\Phi$-derivable approximation defined by truncating $\Phi$ at $O(\lambda)$, Eq.~(\ref{4-point}) can be written diagrammatically as
\begin{equation}
\Phi^{\ts{tr}}[G]=\frac{1}{8}
\spic{
\Oval(15,7.5)(7.5,7.5)(0)
\Oval(15,22.5)(7.5,7.5)(0)
\Vertex(15,15){2}} \ \ \longrightarrow \ \ 
G^{(4,\ts{tr})}\equiv\spic{
\Line(0,0)(15,15)
\Line(0,30)(15,15)
\Line(30,0)(15,15)
\Line(30,30)(15,15)
\GCirc(15,15){6}{0.5}
}=\spic{
\Line(0,0)(15,15)
\Line(0,30)(15,15)
\Line(30,0)(15,15)
\Line(30,30)(15,15)
\Vertex(15,15){2}}
+\half 
\spicc{
\Line(0,0)(15,15)
\Line(0,30)(15,15)
\Curve{(15,15)(30,25)(45,15)}
\Curve{(15,15)(30,5)(45,15)}
\Line(60,0)(45,15)
\Line(60,30)(45,15)
\GCirc(45,15){6}{0.5}
\Vertex(15,15){2}
}
\label{4eq}
\end{equation}
Comparing with the exact Schwinger-Dyson equations of the 4-point function
\begin{equation}
\spic{
\Line(0,0)(15,15)
\Line(0,30)(15,15)
\Line(30,0)(15,15)
\Line(30,30)(15,15)
\GCirc(15,15){6}{0.5}
}=\spic{
\Line(0,0)(15,15)
\Line(0,30)(15,15)
\Line(30,0)(15,15)
\Line(30,30)(15,15)
\Vertex(15,15){2}}
+\half 
\spic{
\Line(0,0)(5,15)
\Line(0,30)(5,15)
\Curve{(5,15)(15,20)(25,15)}
\Curve{(5,15)(15,10)(25,15)}
\Line(30,0)(25,15)
\Line(30,30)(25,15)
\GCirc(25,15){6}{0.5}
\Vertex(5,15){2}}
+\half
\spic{
\Line(0,0)(15,5)
\Line(30,0)(15,5)
\Line(30,30)(15,25)
\Line(0,30)(15,25)
\Oval(15,15)(10,5)(0)
\GCirc(15,5){5}{0.5}
\Vertex(15,25){2}
}
+\half
\spic{
\Line(0,0)(15,5)
\Line(0,30)(15,25)
\Oval(15,15)(10,5)(0)
\Curve{(15,25)(22,20)(25,8)(30,0)}
\Curve{(15,5)(22,10)(25,22)(30,30)}
\SetColor{Black}
\GCirc(15,5){6}{0.5}
\Vertex(15,25){2}
}
+\,{\mbox{2 loops}}
\end{equation}
we see that, at $O(\lambda^2)$, the Bose symmetry of the 4-point function defined from Eq.~(\ref{4eq}) is violated, as only one of the three channels is present. In particular, this implies that at $O(\lambda^2)$, the $\beta$-function in this approximation would be exactly $1/3$ of the perturbative one. This shows that renormalization works quite differently in $\Phi$-derivable approximations\cite{vanhees}.\\
\indent In general, Bose symmetry is guaranteed only up to the order of truncation. The fact that Bose symmetry is violated at higher order has two direct consequences: the renormalization procedure is not trivial (as in the example above), and in gauge theories, Ward identities are generally violated.  

\section{Gauge-fixing dependence}
Let us focus now on gauge theories. Even though Ward identities are violated
for the vertices at higher order than the truncation order, it might happen
that quantities such as the pressure or entropy are gauge invariant. If that
was not the case, we would find a dependence on the gauge condition. For that
reason let us study the dependence of the 2PI-effective action and its truncations on the gauge condition. \\
\indent Consider a pure gauge theory with gauge condition $C[A]$ and parameter $\xi$. The action is given by the usual Yang-Mills part and the gauge-fixing part as
\begin{equation}
S=S_{\ts{YM}}+S_{\ts{GF}}=\int\,-\frac{1}{4} F_{\mu \nu}^{a}F^{\mu \nu}_{a} -\bar{c}^a \frac{\delta C^a}{\delta A_{b\mu}}(D_{\mu}c)_{b}+B_aC^a-\half \xi B_aB^a
\end{equation}
where $A$, $c$ and $\bar{c}$ are the gauge and ghost fields respectively. Having introduced the auxiliary field $B$ permits to write $S_{\ts{GF}}$ as a complete BRST variation. 
\begin{equation}
S_{\tst{\sc GF}}=Q_{\tst{\sc B}}\int \,\half \xi \bar{c}_a B^a-\bar{c}_a C^a\equiv Q_{\tst{\sc B}} \Psi
\end{equation}
Hence, for a gauge condition given by $\Psi$ we can write schematically (using the same symbols indistinctly for all fields)
\begin{equation}
e^{i\Gamma_{\ts{2PI}}[\phi,G]}=\int \mathcal{D} \varphi \, e^{i\left\{S_{\tst{\sc YM}}+Q_{\tst{\sc BRS}}\Psi -J_i (\varphi-\phi)_i+\half K_{ij}(\varphi_i\varphi_j-G_{ij}-\phi_i\phi_j) \right\} }
\end{equation} 
To obtain the dependence of $\Gamma_{\ts{2PI}}$ on the gauge condition $\Psi$ we analyze consistently the change of $\Gamma_{\ts{2PI}}$, $\phi$ and $G$ under the shift $\Psi \rightarrow \Psi + \Delta \Psi$. To achieve this, certain properties of the BRST symmetry are crucial. The details of this procedure are given elsewhere\cite{arrizabalaga}. For small variations $\Delta \Psi$ the result is
\begin{equation}
\Delta \Gamma_{\ts{2PI}}[\phi,G]=\half \Big\langle \Delta \Psi \, Q_{\tst{BRST}} \Big( \widetilde{\phi}_i \frac{\delta \Gamma_{\ts{2PI}}}{\delta \phi_i} + \widetilde{G}_{ij}\frac{\delta \Gamma_{\ts{2PI}}}{\delta G_{ij}}\Big)^2\Big\rangle
\label{result}
\end{equation} 
where $\widetilde{\phi}_i=(\varphi-\phi)_i$, $\widetilde{G}_{ij}=(\varphi-\phi)_i(\varphi-\phi)_j-G_{ij}$ and $Q_{\tst{BRST}}$ is the BRST charge. 
We notice that the 2PI effective action is gauge fixing independent, and thus gauge invariant, at its stationary point.   \\
\indent In a $\Phi$-derivable approximation we are working with a truncated action $\Gamma^{\ts{tr}}_{\ts{2PI}}$. If the truncation is performed at $L$ loops (i.e.~at $O(g^{2L-2})\,$) then $\Gamma_{\ts{2PI}}=\Gamma^{\ts{tr}}_{\ts{2PI}}( \mbox{of $O(g^{2L-2})$})+\Gamma^{\ts{rest}}_{\ts{2PI}}( \mbox{of $O(g^{2L})$})$. Applying Eq.(\ref{result}) and restricting to the physical fields of $\Gamma^{\ts{tr}}_{\ts{2PI}}$ defined by Eq.(\ref{stationarity}) we obtain
\begin{equation}
\Big(\Delta \Gamma^{\ts{tr}}+\Delta \Gamma^{\ts{rest}}\Big)\Big|_{\ts{ph}}=\half \Big\langle \Delta \Psi \, Q_{\tst{BRST}} \Big( \widetilde{\phi}_i \frac{\delta \Gamma^{\ts{rest}}}{\delta \phi_i}\Big|_{\ts{ph}} + \widetilde{G}_{ij}\frac{\delta \Gamma^{\ts{rest}}}{\delta G_{ij}}\Big|_{\ts{ph}}\Big)^2\Big\rangle
\end{equation}
We notice that $\Delta \Gamma^{\ts{tr}} \sim O(\Gamma^{\ts{rest}})\sim O(g^{2L})$,
thus we conclude that $\Gamma^{\ts{tr}}$ is gauge-fixing independent up to the
truncation order. Moreover, at the physical quantities defined from $\Gamma^{\ts{tr}}$, the complete 2PI-effective action $\Gamma$ is gauge-fixing independent up to twice the truncation order since $\Delta \Gamma \sim O([\Gamma^{\ts{rest}}]^2)$. The gauge dependent terms are therefore high order effects controlled by the approximation.   \\
\indent Still, one could take a completely different gauge condition, say
increase $\xi$ by a large amount, so that the gauge dependent terms become
important. However, this should not be done as $\Phi$-derivable approximations
restrict implicitly the choices of gauges available. The point is that the
choice of gauge must not upset the ordering principle that was used in the
approximation (in our case $\Gamma^{\ts{tr}} \gg \Gamma^{\ts{rest}}$). This
fact is familiar from perturbation theory. There, the ordering principle is
$g\ll 1$. Consider the covariant gauge given by $C^a[A]=\partial^{\mu}A^{a}_{\mu}$. The
longitudinal part of the propagator is then proportional to $\xi$. Hence, a
diagram with $I$ internal lines, $L$ loops and $V_{(3,4)}$ vertices has a part
proportional to $(g\xi)^{2L-2}\xi^{V_3/2}$. Taking $|\xi|>1/g$ would upset the
ordering of the diagrams thus ruining perturbation theory. In $\Phi$-derivable
approximations the explicit dependence of $G$ with respect to $\xi$ is in
principle hard to compute, so the same analysis cannot be performed. However,
if we want the approximation to be close to the original path integral, $\xi$
may not be taken far from $\xi \sim 1$, as this would invalidate\cite{arrizabalaga} the use of
the bare coupling constant $g$ for the ordering $\Gamma^{\ts{tr}} \gg
\Gamma^{\ts{rest}}$. \\
\indent We conclude that the error introduced by gauge dependent terms in $\Phi$-derivable approximations is controlled. Improving the approximation would reduce the gauge dependence. Whether this error is smaller than the one introduced by the approximation itself is not known. An explicit $\Phi$-derivable approximation in gauge theories should be carried out. We are confident that the method should be applicable to gauge theories.    
\indent {\bf Acknowledgements:} This work is supported by FOM/NWO.

\end{document}